\documentclass[aip,reprint]{revtex4-1}

\usepackage[utf8]{inputenc}

\usepackage{amsmath}
\usepackage{amssymb}
\usepackage{xspace}
\usepackage{graphicx}
\usepackage{braket}
\usepackage{siunitx}
\usepackage[version=3]{mhchem}
\usepackage{cleveref}

\newcommand{\density}[2]{\gamma_{#2}^{#1}}
\newcommand{\cumulant}[2]{\lambda_{#2}^{#1}}
\newcommand{\cop}[1]{\hat{a}^\dag_{#1}}
\newcommand{\aop}[1]{\hat{a}^{\phantom \dag}_{#1}}
\newcommand{\nname}[0]{$\boldsymbol\lambda_2$-norm\xspace}

\begin{document}

\title{Mutual Correlation}

\author{Francesco A. Evangelista}
\email{francesco.evangelista@emory.edu}
\affiliation{Department of Chemistry and Cherry Emerson Center for Scientific Computation, Emory University, Atlanta, GA 30322, USA}

\crefname{equation}{eq}{eqs}
\creflabelformat{equation}{#2\textup{#1}#3}

\begin{abstract}
Quantifying correlation and complexity in quantum many-body states is central to advancing theoretical and computational chemistry, physics, and quantum information science. This work introduces a novel framework, \textit{mutual correlation}, based on the Frobenius norm squared of the two-body reduced density matrix cumulant.
Through systematic partitioning of the cumulant norm, mutual correlation quantifies nonadditive correlations among interacting subsystems.
Benchmark studies on model systems, including \ce{H10}, \ce{N2}, and \textit{p}-benzyne, demonstrate its efficacy and computational advantage compared to entropy-based metrics such as orbital mutual information.
Maximally correlated orbitals, obtained by maximizing a nonlinear cost function of the mutual correlation, are also considered to identify a basis-independent partitioning of correlation.
This study suggests that mutual correlation is a broadly applicable metric, useful in active space selection and the interpretation of electronic states.
\end{abstract}

\maketitle

\section{Introduction}

Quantifying the complexity of quantum many-body states, whether stationary or evolving in time, is relevant to many aspects of physics, chemistry, and quantum information science.
Measures of complexity find use in the characterization of quantum phase transitions\cite{Osterloh.2002.10.1038/416608a,Osborne.2002.10.1103/physreva.66.032110} and topological order,\cite{Kitaev.2006.10.1103/PhysRevLett.96.110404,Haque.2007.10.1103/physrevlett.98.060401} measuring orbital interactions,\cite{Rissler.2006.10.1016/j.chemphys.2005.10.018,Hachmann.2007.10.1063/1.2768362fix,Freitag.2015.10.1039/c4cp05278a,Lampert.2024.10.1063/5.0232316} improving the performance of numerical algorithms,\cite{Legeza.2003.10.1103/physrevb.68.195116fix} estimating the applicability of approximate computational methods,\cite{Jiang.2012.10.1021/ct2006852,Liu.2020.10.1021/acs.jpclett.0c02288,Duan.2020.10.1021/acs.jctc.0c00358} and identifying important orbitals in active space methods.\cite{Stein.2016.10.1021/acs.jctc.6b00156,Khedkar.2019.10.1021/acs.jctc.8b01293}
Focusing on the domain of first-principle computations, one is particularly interested in those metrics of complexity that can be easily computed and provide useful insights into the problem under study.

Complexity in quantum states manifests itself in different forms, and various methods have been suggested to quantify it.
Statistical metrics of correlation based on the $k$-body reduced density matrices ($k$-RDMs) and their connected components ($k$-cumulants)\cite{Kutzelnigg.1999.10.1063/1.478189} play an important role in the theory of superfluids and superconductors,\cite{Penrose.1956.10.1103/PhysRev.104.576,Yang.1962.10.1103/revmodphys.34.694} and in quantum chemistry.\cite{Mcweeny.1967.10.1002/qua.560010641}
In computational studies, metrics based on the 1-RDM and its eigenvalues (natural occupation numbers)\cite{Bartlett.2020.10.1063/5.0029339} as well as the 2-RDM and its connected component (the 2-cumulant)\cite{Mazziotti.1998.10.1016/s0009-2614(98)00470-9,Kutzelnigg.1999.10.1063/1.478189,Kutzelnigg.2003.10.1002/qua.10751,Misiewicz.2021.10.1063/5.0076888} have been considered.\cite{Juhasz.2006.10.1063/1.2378768,Mazziotti.2012.10.1021/cr2000493,Ganoe.2024.10.1039/d4fd00066h}
The one-particle basis that diagonalizes the 1-RDM defines the natural orbitals,\cite{Lowdin.1955.10.1103/physrev.97.1474} which afford a compact representation of a many-body state\cite{Pulay.1988.10.1063/1.454704,Bofill.1989.10.1063/1.455822,Grev.1992.10.1063/1.462574,Bytautas.2003.10.1063/1.1610434,Abrams.2004.10.1016/j.cplett.2004.07.081,Taube.2008.10.1063/1.2902285,Neese.2009.10.1063/1.3086717,Shu.2015.10.1063/1.4905124,Ratini.2024.10.1021/acs.jctc.3c01325} and are useful to interpret calculations.\cite{Reed.1985.10.1063/1.449360,Reed.1985.10.1063/1.449486} 

A second class of metrics is based on quantum information-theoretic concepts.
Entropy measures of entanglement quantify non-classical correlations between the subsystems of a quantum many-body system.
If one partitions the Hilbert space into a system ($A$) and an environment ($B$), then the von Neumann entropy, is defined as $S = - \mathrm{Tr}(\rho_A \log \rho_A)$, where $\rho_A$ is the reduced density matrix of the system.
The entropy $S$ is a measure of uncertainty for the probability distribution of the system when the environment degrees of freedom are traced out.
Generalizations of this quantity include the Rényi entropy,\cite{Rnyi1961OnMO} extensions to multipartite systems,\cite{Coffman.2000.10.1103/PhysRevA.61.052306} and the entanglement spectrum.\cite{Li.2008.10.1103/physrevlett.101.010504}
When the system consists of a single orbital (site), the corresponding entropy is referred to as one-orbital entropy.\cite{Rissler.2006.10.1016/j.chemphys.2005.10.018}
One may generalize this idea to orbital pairs (two-orbital entropy), and from these quantities define a measure of the entanglement between the two orbitals.\cite{Rissler.2006.10.1016/j.chemphys.2005.10.018}
This quantity is often referred to as \textit{orbital mutual information}.
Entanglement-based metrics have found use in characterizing orbital interactions in calculations of molecules and lattice models.\cite{Rissler.2006.10.1016/j.chemphys.2005.10.018,Hachmann.2007.10.1063/1.2768362fix,Legeza.2003.10.1103/physrevb.68.195116fix,Boguslawski.2012.10.1021/jz301319v,Ding.2023.10.1021/acs.jpclett.3c02536,Ding.2023.10.1088/2058-9565/ad00d9,Ding.2023.10.1088/2058-9565/aca4ee,Aliverti-Piuri.2024.10.1039/D4FD00059E,Liao.2024.10.1021/acs.jpclett.4c01105}
In the case of orbital and mutual information theory, a connection can be established to measures based on the $k$-RDMs; for example, the one-orbital entropy is related to certain elements of the 1- and 2-RDMs, and likewise orbital mutual information depends on $k$-RDMs of up to rank four.\cite{Rissler.2006.10.1016/j.chemphys.2005.10.018}
Similarly, entanglement depth has been connected to the eigenvalues of cumulants of the RDMs.\cite{Liu.2025.10.1103/PhysRevX.15.011056}

A third category of metrics uses geometric and information-theoretic concepts to quantify the complexity of quantum states.
Geometric approaches measure complexity by considering the minimum distance of a state with respect to a separable state.\cite{Vedral.1997.10.1103/physrevlett.78.2275,Aliverti-Piuri.2024.10.1039/D4FD00059E}
Differential geometric concepts such as the quantum geometric tensor have also been proposed to measure the complexity of quantum states.\cite{Shimony.1995.10.1111/j.1749-6632.1995.tb39008.x,Barnum.2001.10.1088/0305-4470/34/35/305,Wei.2003.10.1103/PhysRevA.68.042307,Nico-Katz.2023.10.1103/PhysRevResearch.5.013041}
Another class of approaches is based on the circuit model of quantum computing, where the distance is measured as the length of a quantum circuit.
Quantum circuit complexity is then defined as the size of the smallest sequence of quantum gates needed to prepare a given pure state starting from a separable state.
\cite{Nielsen.2006.10.1126/science.1121541}
Another way to characterize quantum complexity is \textit{magic}, which distinguishes quantum operations into Clifford (classical computable) and non-Clifford operations.
The magic resource needed to prepare a state is the number of non-Clifford operations.
As mentioned above, these metrics quantify different aspects of the complexity of a quantum state.
For example, quantum circuit complexity behaves differently than entanglement.\cite{Aaronson.2016.10.48550/arXiv.1607.05256}
Numerical studies on model systems have also highlighted differences in metrics of correlation.\cite{Stair.2020.10.1063/5.0014928,Ganoe.2024.10.1039/d4fd00066h}

In numerical applications, one is often interested in estimating the degree of correlation or the reliability of a computation based on metrics of complexity.\cite{Jiang.2012.10.1021/ct2006852,Xu.2025.10.1063/5.0250636}
Several diagnostics have been developed to assess the quality of configuration interaction\cite{Coe.2014.10.1016/j.cplett.2014.04.050,Coe.2015.10.1021/acs.jctc.5b00543,Izsak.2023.10.1021/acs.jctc.3c00122} and coupled cluster wave functions.\cite{Lee.1989.10.1002/qua.560360824,Janssen.1998.10.1016/S0009-2614(98)00504-1,Nielsen.1999.10.1016/s0009-2614(99)00770-8}
These diagnostics have also been used to construct machine learning models that can be used to predict wave function-based diagnostics for high-throughput applications.\cite{Liu.2020.10.1021/acs.jpclett.0c02288,Duan.2020.10.1021/acs.jctc.0c00358}

In this work, we consider metrics of correlation based on the 2-RDM cumulant $\boldsymbol{\lambda_2}$ (2-cumulant), specifically the square Frobenius norm ($\| \boldsymbol{\lambda_2}\|_\mathrm{F}^2$).
Juh\'{a}sz and Mazziotti\cite{Juhasz.2006.10.1063/1.2378768} first proposed the use of this quantity as a measure of correlation, highlighting several of its advantages.
These include the additivity of $\| \boldsymbol{\lambda_2}\|_\mathrm{F}^2$ for separable states of noninteracting fragments and invariance under unitary transformations of the one-particle basis.
Moreover, $\| \boldsymbol{\lambda_2}\|_\mathrm{F}^2$ also measures spin entanglement, yielding a nonadditive contribution for spin-coupled noninteracting systems.
This metric of correlation has been examined numerically,\cite{Stair.2020.10.1063/5.0014928} including in a recent analysis by Ganoe and Shee,\cite{Ganoe.2024.10.1039/d4fd00066h} which highlighted its usefulness in quantifying correlation in different regimes.
Alcoba and co-workers proposed a functional of $\boldsymbol{\lambda_2}$ which quantifies electron correlation and enables the detection of spin entanglement.\cite{Alcoba.doi:10.1063/1.3503766}
Note that since $\boldsymbol{\lambda_2}$ grows with system size, it is not a direct measure of strong correlation, and it is not comparable between systems of different chemical compositions.
Alternative measures of correlation based on the eigenvalues of $\boldsymbol{\lambda_2}$ have been discussed by Raeber and Mazziotti.\cite{Raeber.2015.10.1103/physreva.92.052502}
Like the Frobenius norm squared, these quantities are also invariant with respect to unitary transformations of the basis.
In particular, the largest absolute eigenvalue of $\boldsymbol{\lambda_2}$ ($\lambda_\mathrm{max}$) can serve as a measure of strong correlation. Large values of $\lambda_\mathrm{max}$ signal the emergence of off-diagonal long-range order in the 2-RDM, which corresponds to fermion-pair condensation.\cite{Raeber.2015.10.1103/physreva.92.052502}
An analogous result was obtained for the eigenvalues of the cumulant of the particle-hole RDM, with large eigenvalues signaling exciton condensation.\cite{Schouten.2022.10.1103/physrevb.105.245151}
The maximal eigenvalues of $\boldsymbol{\lambda_2}$ based on a different reshaping of $\boldsymbol{\lambda_2}$ as a matrix has been considered more recently in the context of entanglement witnessing using x-ray spectroscopy.\cite{Liu.2025.10.1103/PhysRevX.15.011056}
This metric provides a lower bound for the entanglement depth, defined as the size of the largest cluster of electrons whose state cannot be factorized into a product.

This paper introduces metrics of correlations obtained by partitioning $\| \boldsymbol{\lambda_2}\|_\mathrm{F}^2$ into contributions from subsets of the orbital basis (fragments), in a way analogous to many-body decompositions of the energy.\cite{Gordon.2012.10.1021/cr200093j}
This partitioning enables the quantification of nonadditive correlation that arises from interacting parts of quantum systems, which we refer to as \textit{mutual correlation}.
We show that mutual correlation is a useful metric for qualitative analysis of electronic structure, serving as an economical alternative to established metrics based on entropy.
To this end, we show that mutual correlation is applicable to problems of active space selection, and it lends itself to defining orbitals that minimize or localize mutual correlation among a limited set of orbitals.

A practical advantage of formulating a metric of correlation based on the 2-body cumulant is the ease with which it can be computed from the 2-RDM, a quantity widely available in many software implementations and for a variety of many-body methods and measurable via efficient quantum algorithms.\cite{Zhao.2021.10.1103/physrevlett.127.110504}
In contrast, orbital mutual information requires higher-order RDMs (up to a quadratic number of elements of the 4-RDM) and so far it has been applied extensively only in conjunction with the density matrix renormalization group (DMRG).\cite{White.1992.10.1103/physrevlett.69.2863}
We note that to obviate this limitation, Liao, Ding, and Schilling\cite{Liao.2024.10.1021/acs.jpclett.4c01105} have suggested using instead the total orbital correlation, defined as the sum of single orbital entropies.

The paper is organized as follows.
In \cref{sec:theory} we review the definition of reduced density matrices and the corresponding correlation measures. We also discuss a partitioning of the total correlation and define mutual correlation.
In \cref{sec:results}, we quantify orbital interactions by constructing a (pairwise) orbital mutual correlation measure.
We calibrate this measure of correlation and apply it to several representative molecules with various correlation strengths.
This section also considers orbital transformations that extremize the orbital mutual correlation and a comparison with orbital entanglement metrics.
In \cref{sec:conclusion}, we summarize our findings and consider future extensions of this work.

\section{Theory}
\label{sec:theory}

\subsection{Reduced density cumulants and correlation metrics}

Consider a normalized $N$-particle quantum state $\ket{\Psi}$ expanded in an orthonormal one-particle spin orbital basis of dimension $2K$, $ S = \{ \ket{\psi_{p}} \}_{p=1}^{2K}$.
We consider an unrestricted spin orbital basis, where each element is the product of a spin-dependent (spatial) orbital $\ket{\phi_P^{\sigma}}$ and a spin function $\ket{\sigma} \in \{\ket{\uparrow}, \ket{\downarrow} \}$:
\begin{equation}
\ket{\psi_{p}} \equiv \ket{\psi_{P\sigma}} = \ket{\phi_P^{\sigma}} \otimes \ket{\sigma}
\end{equation}
The orbitals $\{ \ket{\phi_P^{\sigma}} \}$ span a space of dimension $K$.
For convenience, the spin orbitals are labeled by a lowercase composite index ($p$) that combines the orbital and spin indices, $p \equiv (P,\sigma)$.

The one- and two-body reduced density matrices ($\boldsymbol \gamma_1$ and $\boldsymbol \gamma_2$) are defined respectively as:
\begin{equation}
\density{p}{r} = \braket{\Psi | \cop{p} \aop{r}|\Psi}
\end{equation}
and
\begin{equation}
\density{pq}{rs} = \braket{\Psi | \cop{p} \cop{q} \aop{s} \aop{r} |\Psi}
\end{equation}
where $\cop{p}$ ($\aop{p}$) is a fermionic second quantization creation (annihilation) operator (see Refs.~\citenum{Kutzelnigg.1982.10.1063/1.444231} and \citenum{Kutzelnigg.1997.10.1063/1.474405} for the notation used here).
Defined this way, the 1- and 2-RDMs satisfy the following trace relationships:
\begin{equation}
\begin{split}
\mathrm{Tr}(\boldsymbol{\gamma_1}) & = \sum_{p} \density{p}{p} = N \\ 
\mathrm{Tr}(\boldsymbol{\gamma_2}) & = \sum_{pq} \density{pq}{pq} = N (N-1)
\end{split}
\end{equation}

The 2-cumulant ($\boldsymbol{\lambda_2}$) is the connected (extensive) part of the two-body reduced density matrix.
This quantity captures genuine two-particle correlations that do not reduce to products of 1-RDMs, and may be expressed in terms of $\boldsymbol \gamma_1$ and $\boldsymbol \gamma_2$ as:
\begin{equation}
\label{eq:2cdm_def}
\cumulant{pq}{rs} = \braket{\Psi | \cop{p} \cop{q} \aop{s} \aop{r} |\Psi}_\mathrm{c} = \density{pq}{rs} - (\density{p}{r} \density{q}{s} - \density{p}{s} \density{q}{r})
\end{equation}
Both $\density{pq}{rs}$ and $\cumulant{pq}{rs}$ are antisymmetric with respect to distinct permutations of upper and lower indices ($\density{pq}{rs} = -\density{qp}{rs} = -\density{pq}{sr} = \density{qp}{sr}$), and Hermitian ($\density{pq}{rs}= {\density{rs}{pq}}^*$).

In this work, we are concerned with quantifying correlation with the square of the Frobenius norm of the 2-cumulant:\cite{Juhasz.2006.10.1063/1.2378768} 
\begin{equation}
\mathcal{C} \equiv \frac{1}{4} \| \boldsymbol{\lambda_2}\|_\mathrm{F}^2 = \frac{1}{4}\sum_{pqrs}^{2K} |\cumulant{pq}{rs}|^2
\end{equation}
For convenience, we denote this quantity as $\mathcal{C}$ and refer to it as \nname correlation (\nname).
Note that our definition of $\mathcal{C}$ includes a factor of one-fourth to account for repeated terms in the sum.
This metric of correlation has several desirable properties.
$\mathcal{C}$ is null for a Slater determinant (since $\boldsymbol{\lambda}_2 = 0$) and additive for a state separable into noninteracting fragments in a localized basis.
Furthermore, $\mathcal{C}$ is invariant with respect to unitary transformations of the orbital basis.
Combining these facts, $\mathcal{C}$ is additive for separable states irrespective of the spin orbital basis used.
Moreover, $\mathcal{C}$ is null for one-electron systems.

Alternative metrics of correlation based on other properties of the cumulants have been proposed.
One such metric is the trace of the 2-cumulant, which appears in theories that attempt to reconstruct the 1-RDM from the 2-cumulant.\cite{Kutzelnigg.1999.10.1063/1.478189,Kutzelnigg.2006.10.1063/1.2387955,Misiewicz.2020.10.1063/5.0036512}
The trace of the 2-cumulant is related to the deviation from idempotency of the 1-RDM:
\begin{equation}
\begin{split}
\mathrm{Tr}(\boldsymbol{\lambda}_2) & =
 \sum_{pq} \cumulant{pq}{pq} = 
 \mathrm{Tr}(\boldsymbol{\gamma}_2) - [\mathrm{Tr}(\boldsymbol{\gamma}_1)]^2 + 
 \sum_{pq} \density{p}{q} \density{q}{p} \\
& = -N + \sum_{pq} \density{p}{q} \density{q}{p} =  \mathrm{Tr}(\boldsymbol{\gamma}_1^2 - \boldsymbol{\gamma}_1)
\end{split}
\end{equation}
The dependence of this metric on the simpler 1-RDM suggests that it does not account for all correlations encoded in the 2-cumulant.
Furthermore, Kong and Valeev have argued\cite{Kong.2011.doi:10.1063/1.3596948} that the diagonal elements of the 2-cumulant should not be considered viable indicators of the amount of correlation independently of the basis in which it is represented.

Metrics based on the eigenvalues of $\boldsymbol{\lambda}_2$ have also been proposed.\cite{Juhasz.2006.10.1063/1.2378768,Raeber.2015.10.1103/physreva.92.052502}
One metric corresponds to reshaping $\boldsymbol{\lambda}_2$ into a matrix with indices that correspond to pairs of creation and annihilation operators separately. We refer to this as the geminal matricization, which here we define as:
\begin{equation}
\label{eq:Mgem}
(\mathbf{M}^\mathrm{gem})_{[pq],[rs]} = \cumulant{pq}{rs} \quad p < q, r < s
\end{equation}
Note that in defining $\mathbf{M}^\mathrm{gem}$, the composite spinorbital indices $[pq]$ and $[rs]$ run over only the unique pairs ($2K^2 - K$).
One can alternatively include all pairs, but the eigenvalues and eigenvectors will be numerically different.
Alternatively, one may reshape $\boldsymbol{\lambda}_2$ according to the pairs of creation and annihilation indices to obtain a particle-hole (ph) matricized version:
\begin{equation}
\label{eq:Mph}
(\mathbf{M}^\mathrm{ph})_{[pq],[rs]} = \cumulant{pr}{qs} \quad \forall pq, \forall rs
\end{equation}

\begin{figure}[htbp!]
    \includegraphics[width=3in]{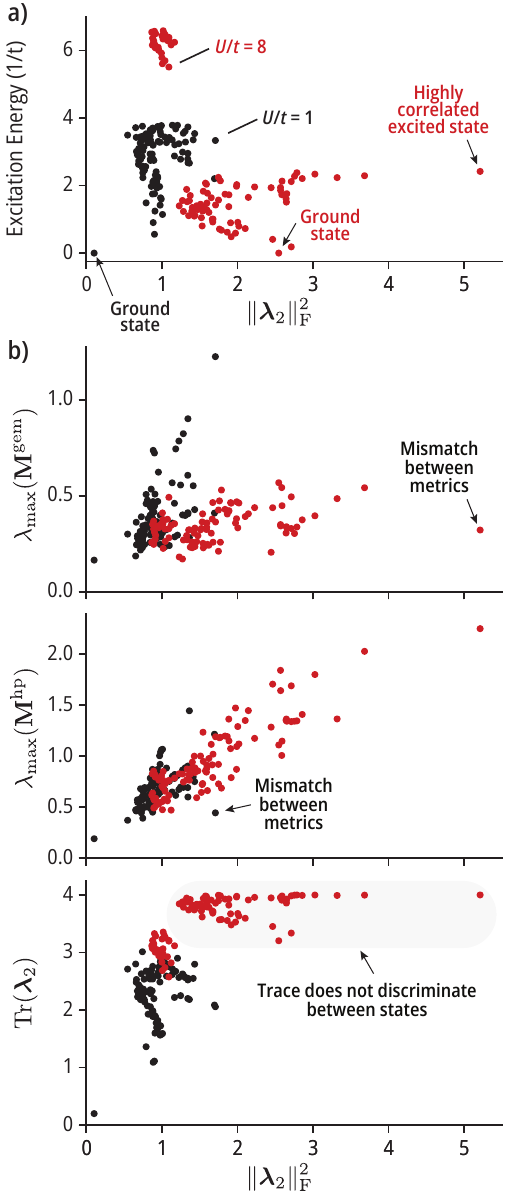}
    \caption{One-dimensional Hubbard model with eight sites at half filling (4 spin up and 4 spin down electrons).
    Analysis of the lowest 100 states for the cases $U/t$ = 1 (black dots) and 8 (red dots).
        a) Plot of the excitation energy (in units of $U/t$) vs. the Frobenius norm squared ($\| \boldsymbol{\lambda}_2 \|_\mathrm{F}^2$) of the 2-cumulant.
        b) Distribution of the values of the largest absolute eigenvalue of the $\boldsymbol{\lambda}_2$ from $\mathbf{M}^\mathrm{gem}$ [top, see \cref{eq:Mgem}] and $\mathbf{M}^\mathrm{ph}$ [middle, see \cref{eq:Mph}] and the trace of $\boldsymbol{\lambda}_2$ [bottom, $\mathrm{Tr}(\boldsymbol{\lambda}_2)$].}
    \label{fig:hubbard_8sites}    
\end{figure}

To appreciate the difference between metrics of correlation based on $\boldsymbol{\lambda}_2$, we examine the eigenstates of the one-dimensional Hubbard model Hamiltonian with open boundary conditions:
\begin{equation}
H = - t \sum_{p=1}^{K-1} \sum_{\sigma\in\{\uparrow,\downarrow\}} (\cop{p + 1,\sigma} \aop{p,\sigma} + \cop{p,\sigma} \aop{p+1,\sigma}) + U \sum_{p=1}^{K} \hat{n}_{p\uparrow} \hat{n}_{p\downarrow}
\end{equation}
Here $\hat{n}_{p\sigma} = \cop{p\sigma} \aop{p\sigma}$ is the number operator for the $p$-th site and $K$ is the total number of sites.
We first focus on a system with eight sites at half filling (4 spin up and 4 spin down electrons).
In \cref{fig:hubbard_8sites} a), we show the distribution of excitation energies (computed with respect to the ground state) and values of the square of the Frobenius norm for the lowest 100 states, in the case of a small  ($U/t$ = 1) and large  ($U/t$ = 8) ratio between the on-site repulsion ($U$) and hopping ($t$) parameters.
For the case $U/t$ = 1, the square of the Frobenius norm characterizes the ground state as being the least correlated, while for $U/t$ = 8, the ground state is characterized by a higher degree of correlation than all the lowest 100 states at $U/t$ = 1.
Numerical maximization of the Frobenius norm squared over all states in the Hilbert space of 4 spin up and 4 spin down electrons distributed among 8 sites leads to the upper bound $\|\boldsymbol{\lambda}_2\|_\mathrm{F}^2 < 7.5$.
Therefore, the ground state of this Hamiltonian for both parameter values explored is far from representing a state that maximizes this correlation metric.

In \cref{fig:hubbard_8sites} b), we show the correlation between other correlation metrics based on $\boldsymbol{\lambda}_2$ and $\|\boldsymbol{\lambda}_2\|_\mathrm{F}^2$ for the first 100 states of the Hubbard Hamiltonian.
Among these metrics, $\lambda_\mathrm{max}(\mathbf{M}^\mathrm{ph})$ is the one that shows the best correlation with the Frobenius norm-based metric.
For the $\lambda_\mathrm{max}(\mathbf{M}^\mathrm{gem})$ metric, the correlation is less consistent, and one finds false negatives, that is, states that would be characterized by this metric as weakly-to-medium correlated (see \cref{fig:hubbard_8sites}), which the other three metrics identify as the state with the largest value degree of correlation.
Lastly, the trace of $\boldsymbol{\lambda}_2$ appears to be less capable of discriminating correlation in the states of the Hubbard Hamiltonian when $U/t$ = 8.

\begin{table*}
\caption{Examples of four electron states (two spin up and two spin down) in a four orbital basis that maximize the value of $\| \boldsymbol{\lambda}_2 \|_\mathrm{F}^2$ together with the values of the largest absolute eigenvalue of the $\boldsymbol{\lambda}_2$ from $\mathbf{M}^\mathrm{gem}$ [\cref{eq:Mgem}] and $\mathbf{M}^\mathrm{ph}$ [\cref{eq:Mph}] and the trace of $\boldsymbol{\lambda}_2$ [$\mathrm{Tr}(\boldsymbol{\lambda}_2)$].
The bottom row shows the maximum value of each metric of correlation.
We use the symbols $(0,\uparrow,\downarrow,2\equiv\uparrow\downarrow)$ to represent the occupation of each orbital.
Values in bold indicate large discrepancies among the metrics of correlation based on $\boldsymbol{\lambda}_2$.
}
\footnotesize
\renewcommand{\arraystretch}{1.75}
\begin{tabular}{lcccc}
\hline

\hline
State & $\| \boldsymbol{\lambda}_2 \|_\mathrm{F}^2$ & $|\mathrm{Tr}(\boldsymbol{\lambda}_2)|$ & $\lambda_\mathrm{max}(\mathbf{M}^\mathrm{gem})$ & $\lambda_\mathrm{max}(\mathbf{M}^\mathrm{ph})$ \\
\hline 
$\ket{\Psi_1} = \displaystyle \frac{1}{\sqrt{2}}
\left(
\ket{\uparrow \downarrow \downarrow \uparrow} - \ket{ \downarrow \uparrow \uparrow \downarrow}
\right)$
& 1.750 & 2.000 & \textbf{0.250} & 1.750 \\
$\ket{\Psi_2} =\displaystyle \frac{1}{\sqrt{2}}
\left(
\ket{0 2 2 0} - \ket{2 0 0 2}
\right)$
& 1.750 & 2.000 & \textbf{0.250} & 1.750 \\
$\ket{\Psi_3} =\displaystyle \frac{1}{2 \sqrt{2 + \sqrt{2}}}
\left(
\ket{0022} - \ket{2200} + (1 + \sqrt{2}) (\ket{0202} - \ket{2020})
\right)
$
& 1.750 & 2.000 & \textbf{0.604} & 1.457
\\
$\ket{\Psi_4} =\displaystyle \frac{1}{\sqrt{6}}
\left(
\ket{2200} + \ket{2002} + \ket{0220} + \ket{0022} - \ket{2020} -\ket{0202}
\right)
$
& 1.750 & 2.000 & 1.250 & \textbf{0.417} \\
Maximum value & 1.750 & 2.000 & 1.250 & 1.750 \\
\hline

\hline
\end{tabular}
\label{tab:hubbard_examples}
\end{table*}

To gain more insight from the perspective of the quantum state, in \cref{tab:hubbard_examples} we consider a 4 site model with 2 spin up and 2 spin down electrons.
For this system, we list representative states that maximize $\|\boldsymbol{\lambda}_2\|_\mathrm{F}^2$ but are not characterized as highly correlated by the eigenvalue-based metrics.
For example, the singlet state:
\begin{equation}
\ket{\Psi_2} = \frac{1}{\sqrt{2}} \left(\ket{0220} - \ket{2002} \right)
\end{equation}
is analogous to a Bell state and is not separable.
In accordance, for this state $\|\boldsymbol{\lambda}_2\|_\mathrm{F}^2$, $\mathrm{Tr}(\boldsymbol{\lambda}_2)$, and $\lambda_\mathrm{max}(\mathbf{M}^\mathrm{ph})$ take the maximum value allowed within the given Hilbert space.
In contrast, the $\lambda_\mathrm{max}(\mathbf{M}^\mathrm{gem})$ metric characterizes this state as not maximally correlated, taking a value (0.25) that is only 1/5 of the maximum value.
However, looking beyond only the largest absolute eigenvalue of $\mathbf{M}^\mathrm{gem}$, we find that the eigenvalues are identical and equal to 0.25.
State $\ket{\Psi_4}$ reported in \cref{tab:hubbard_examples} behaves in the opposite way, maximizing $\lambda_\mathrm{max}(\mathbf{M}^\mathrm{gem})$, but yielding a value of $\lambda_\mathrm{max}(\mathbf{M}^\mathrm{ph}) = 0.417$ that is approximately 1/4 of the maximum value for this metric.
Examination of the absolute eigenvalues of $\mathbf{M}^\mathrm{ph}$ for this case shows that instead of a single large eigenvalue, all eigenvalues take the value of 0.417 or 0.250.
This analysis suggests that eigenvalue metrics based on the largest absolute eigenvalue are not always able to detect strongly correlated states.
This comparison strengthens the argument for using $\|\boldsymbol{\lambda}_2\|_\mathrm{F}^2$ as a metric of electron correlation.

\subsection{Partitioning of the correlation metric}
\label{sec:partitioning}

Next, we proceed to define metrics of mutual correlation by decomposing $\mathcal{C}$ into contributions from subsystems of the full quantum system.
An important motivation for decomposing the \nname is that this quantity measures the sum of correlation effects.
As such, it does not enable comparing systems with different compositions or number of particles.
Moreover, it does not provide a way to discern the distribution of correlation effects.
For example, a certain value of $\mathcal{C}$ may arise due to a small cluster of highly correlated particles or a medium strength between many particles.

We first consider a simple bipartition of the spin orbital basis into subsets $A$ and $B$ ($A \cup B = S$, $A \cap B = \emptyset$).
For each subsystem $X = A$ or $B$, we can define the corresponding \nname correlation:
\begin{equation}
\mathcal{C}_X = \frac{1}{4}\sum_{pqrs \in X} |\cumulant{pq}{rs}|^2
\end{equation}
We define the mutual correlation between subsystems $A$ and $B$ ($\mathcal{M}_{AB}$) as the difference between the total \nname ($\mathcal{C}_{S}$) and the contribution of each subsystem:
\begin{equation}
\mathcal{M}_{AB} = \mathcal{C}_{S} - \mathcal{C}_A - \mathcal{C}_B
\end{equation}
Note that $\mathcal{M}_{AB}$ is positive semidefinite ($\mathcal{M}_{AB} \geq 0$) and equal to zero if the state is separable into a product of states for subsystems $A$ and $B$.

To understand the behavior of $\mathcal{M}_{AB}$, we consider a toy model consisting of two electrons in two spatial orbitals of different symmetry $\ket{\phi_1}$ and $\ket{\phi_2}$.
The behavior of the 2-cumulant for a more complex version of this toy model that does not assume symmetry restrictions has been discussed in Ref.~\citenum{Kong.2011.doi:10.1063/1.3596948}.
For this model, we define the subsets $A$ and $B$ to span the spin orbitals built from $\ket{\phi_1}$ and $\ket{\phi_2}$:
\begin{equation}
\begin{split}
A = \{  \ket{\phi_1}\otimes \ket{\uparrow}, \ket{\phi_1}\otimes \ket{\downarrow} \} \\
B = \{ \ket{\phi_2}\otimes \ket{\uparrow}, \ket{\phi_2}\otimes \ket{\downarrow} \}
\end{split}
\end{equation}
A general singlet, totally symmetric state can be written as a linear combination of two determinants:
\begin{equation}
\ket{\Psi} = \cos \theta \ket{20} + \sin \theta \ket{02}
\end{equation}
where we represent the orbital occupation of an orbital with the values $\{0, \uparrow, \downarrow, 2 \equiv \uparrow \downarrow\}$.
For this model, the values of $\mathcal{C}_A$ and $\mathcal{C}_B$ are identical (since $\lambda^{ {1 \uparrow} {1 \downarrow} }_{ {1 \uparrow} {1 \downarrow} } = \lambda^{ {2 \uparrow} {2 \downarrow} }_{ {2 \uparrow} {2 \downarrow} }$) and are given by:
\begin{equation}
\mathcal{C}_A = \mathcal{C}_B = |\lambda^{ {1 \uparrow} {1 \downarrow} }_{ {1 \uparrow} {1 \downarrow} }|^2 = \cos^4(\theta) \sin^4(\theta)
\end{equation}
$\mathcal{C}_{A}$ is zero when $\ket{\Psi}$ is a single determinant ($\theta = k \pi$, with $k \in \mathbb{Z}$), signaling absence of correlation.
The \nname correlation of $A$ instead assumes the largest value ($1/16$) for $\theta = \pi/4 + k \pi$, corresponding to entangled (Bell) states of the form:
\begin{equation}
\ket{\Psi} = \frac{\pm 1}{\sqrt{2}} (\ket{20} \pm \ket{02})
\end{equation}
The mutual correlation between orbitals 1 and 2 is given by:
\begin{equation}
\mathcal{M}_{AB} = (1/8) [5 - \cos(4 \theta)] \sin^2(2 \theta)
\end{equation}
The minima and maxima of this quantity coincide with those of $\mathcal{C}_{A}$, with values of $\mathcal{M}_{AB}$ ranging from zero (single Slater determinant) to 3/4 (maximum entanglement).
Combining these results, we find that the maximum value of $\mathcal{C}_S$ is 7/8.
 
\subsection{General partitioning}
Bipartite mutual correlation may be extended to a general partitioning of the spin orbital basis $S$ into $n$ disjoint sets $\{ A_k \}_{k=1}^{n}$:
\begin{equation}
	S = \cup_{k=1}^{n} A_k, \quad A_i \cap A_j = \emptyset \quad \forall i,j \in \{1,\cdots,n\}
\end{equation}
These subsets represent a formal partitioning that allows for the analysis of correlations within the chosen basis and need not represent physical subsystems of the full system.
In the general case, the total \nname can be partitioned into contributions from individual subsystems ($\mathcal{C}_{A}$), plus quantities that capture the mutual correlation among two ($\mathcal{M}_{AB}$), three ($\mathcal{M}_{ABC}$), and four ($\mathcal{M}_{ABCD}$) distinct subsystems (where $A,B,C,D \in \{ A_k \}_{k=1}^{n}$):
\begin{equation}
\begin{split}
\mathcal{C}_{S} =& \sum_A \mathcal{C}_{A}
+ \sum_{A<B} \mathcal{M}_{AB} + \sum_{A<B<C} \mathcal{M}_{ABC}
\\
&
+ \sum_{A<B<C<D} \mathcal{M}_{ABCD}
\end{split}
\end{equation}
The mutual correlation between two subsystems may be expressed in terms of the 2-body cumulants as:
\begin{equation}
\label{eq:M_AB}
\begin{split}
\mathcal{M}_{AB} = &
\sum_{p \in A} \sum_{qrs \in B} |\cumulant{pq}{rs}|^2 
+ \frac{1}{2} \sum_{pq \in A} \sum_{rs \in B} |\cumulant{pq}{rs}|^2 \\
& + \sum_{pr \in A} \sum_{qs \in B} |\cumulant{pq}{rs}|^2
+ \sum_{pqr \in A} \sum_{s \in B} |\cumulant{pq}{rs}|^2
\end{split}
\end{equation}
where we have exploited the hermiticity of the cumulant to reduce the number of summations.
Similarly, $\mathcal{M}_{ABC}$ and $\mathcal{M}_{ABCD}$ can be written compactly as:
\begin{equation}
\begin{split}
\mathcal{M}_{ABC} = &
\sum_{p \in A} \sum_{q \in B} \sum_{rs \in C} (|\cumulant{pq}{rs}|^2 + 2 |\cumulant{pr}{qs}|^2) \\
&+ \sum_{p \in B} \sum_{q \in C} \sum_{rs \in A} (|\cumulant{pq}{rs}|^2 + 2 |\cumulant{pr}{qs}|^2) \\
&+ \sum_{p \in A} \sum_{q \in C} \sum_{rs \in B} (|\cumulant{pq}{rs}|^2 + 2 |\cumulant{pr}{qs}|^2)
\end{split}
\end{equation}
and
\begin{equation}
\mathcal{M}_{ABCD} = 6 
\sum_{p \in A} \sum_{q \in B} \sum_{r \in C} \sum_{s \in D} |\cumulant{pq}{rs}|^2
\end{equation}
The numerical value of these mutual correlation metrics depends on the chosen basis, but each term is invariant with respect to unitary transformations that mix spinorbitals within the same subsystems.
Note that while the total correlation $\mathcal{C}_S$ grows with system size, the mutual correlation contributions $\mathcal{M}_{AB}$, $\mathcal{M}_{ABC}$, and $\mathcal{M}_{ABCD}$ are intensive quantities.
Therefore, they characterize intrinsic correlation effects and can be directly compared among different systems.

It would be highly desirable to understand the behaviour and properties of these three classes of mutual correlation.
However, for the purpose of this work, we limit our analysis to the mutual correlation between two subsystems ($\mathcal{M}_{AB}$), leaving the analysis of the more complex terms for future studies.

\newpage

\section{Results}
\label{sec:results}

\subsection{Orbital mutual correlation}

\begin{figure}[hbt!]
    \includegraphics[width=3in]{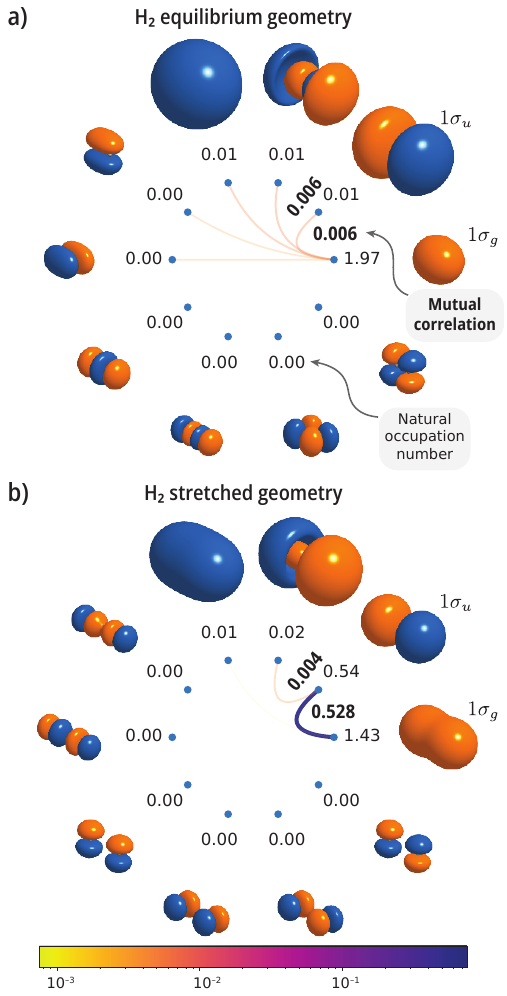}
    \caption{Singlet ground state of the \ce{H2} molecule. Mutual correlation plots computed with an active space comprising the 1s, 2s, and 2p shells [CASSCF(2,10)] using a cc-pVDZ basis set at (a) the experimental geometry ($r_e$ = \SI{0.74144}{\angstrom}), and (b) at a stretched geometry ($r_{\ce{H-H}} = 3r_e$). The numbers next to each orbital show its occupation number, while the color scale on the right corresponds to pairwise mutual correlation.}
    \label{fig:h2}    
\end{figure}

To analyze orbital interactions in molecules, we specialize the pair mutual correlation $\mathcal{M}_{AB}$ to the case of a restricted basis and define the fragments as
\begin{equation}
A_P = \{ \ket{\phi_{P}} \otimes \ket{\uparrow}, \ket{\phi_{P}} \otimes \ket{\downarrow} \}
\end{equation}
With this partitioning, the mutual correlation between orbital pairs $P$ and $Q$ [$\mathcal{M}_{P Q}$, see \cref{eq:M_AB}] may be expressed as the sum of spin-free quantities:
\begin{equation}
\begin{split}
\mathcal{M}_{P Q} = {(\Lambda^2)}^{PQ}_{QQ}
+ \frac{1}{2} {(\Lambda^2)}^{PP}_{QQ}
+ {(\Lambda^2)}^{PQ}_{PQ}
+ {(\Lambda^2)}^{PP}_{QP}
\end{split}
\end{equation}
where ${\Lambda^2}^{PQ}_{RS}$ is a sum of squares of elements of the two-body cumulant:
\begin{equation}
\begin{split}
{(\Lambda^2)}^{PQ}_{RS} = 
& |\cumulant{P\uparrow Q\uparrow}{R\uparrow S\uparrow}|^2
+ |\cumulant{P\uparrow Q\downarrow}{R\uparrow S\downarrow}|^2
+ |\cumulant{Q\uparrow P\downarrow}{R\uparrow S\downarrow}|^2 \\
&  + |\cumulant{P\uparrow Q\downarrow}{R\uparrow S\downarrow}|^2
+ |\cumulant{Q\uparrow P\downarrow}{R\uparrow S\downarrow}|^2
+ |\cumulant{P\downarrow Q\downarrow}{R\downarrow S\downarrow}|^2
\end{split}
\end{equation}
Here, we assume that $\ket{\Psi}$ is an eigenvector of $\hat{S}_z$, so we consider only those terms that preserve the corresponding quantum number $M_S$.
As found for the two-orbital toy model discussed in \cref{sec:partitioning}, the value of $\mathcal{M}_{P Q}$ ranges from 0 to 3/4.

\subsection{Visualization and calibration}

Having defined orbital mutual correlation, we establish its sensitivity with respect to the basis set, propose a way to visualize it, and develop a way to interpret its numerical value. 
In the following results, we obtain the 2-cumulant from a CASSCF computation using the cc-pVDZ basis set (except when noted) and various active spaces.
After convergence, we transform the orbitals to the natural basis in which the spin-summed 1-RDM ($\density{P\uparrow}{Q\uparrow} + \density{P\downarrow}{Q\downarrow}$) is diagonal.
The diagonal elements of the 1-RDM then correspond to the natural orbital occupation numbers (NOONs), defined as $n_{P} = \density{P\uparrow}{P\uparrow} + \density{P\downarrow}{P\downarrow}$.
Note that the natural basis is unique, up to degeneracies in the occupation number.
All computations were performed using the Forte\cite{Evangelista.2024.10.1063/5.0216512} software package using molecular integrals obtained from Psi4.\cite{Smith.2020.10.1063/5.0006002}
Mutual correlation plots were made with the VMD program\cite{HUMP96} using the VMDCube Python package.\cite{vmdcube}

A criterion for considering a metric of correlation useful is a weak dependence with respect to the computational basis (for a sufficiently large basis).
For this purpose, we compute total and mutual correlation metrics for the nitrogen molecule using a full valence active space comprising the 2s and 2p orbitals and correlating 8 electrons.
We consider both the equilibrium ($r_e$) and a stretched geometry ($2r_e$) to examine both the cases of weak and strong entanglement.
\cref{tab:basis_set} reports $\mathcal{C}$ and the pair mutual correlation ($\mathcal{M}_{AB}$) for basis sets ranging from cc-pVDZ (28 functions) to cc-pVQZ (110 functions),\cite{Dunning.1989.10.1063/1.456153} with the latter representing the typical standard for high-level computations.
These data show very little dependence of both quantities on the basis set, with the largest relative changes being observed at the stretched geometry.
For example, at $2r_e$ going from the cc-pVDZ to the cc-pVQZ, the value of $\mathcal{C}$ changes only by 0.48\%, and changes of a similar or smaller ratio are observed for the mutual correlation.

\begin{table}
\renewcommand{\arraystretch}{1.1}
\caption{Singlet ground state of the nitrogen molecule. Comparison of the total \nname correlation metric [$\mathcal{C}$] and mutual correlation ($\mathcal{M}_{AB}$) as a function of basis set size. Mutual correlations is shown for the four orbital pairs with the largest value. All values are computed at the CASSCF level with a full valence active space [CASSCF(10e,8o)] at bond lengths equal to one and two times the equilibrium distance ($r_e$ = \SI{1.098}{\angstrom}, taken from Ref.~\citenum{Huber.1979.10.1007/978-1-4757-0961-2_2}).
}
\large
\begin{tabular}{lccc}
\hline

\hline
Quantity & & Basis set & \\
  & cc-pVDZ & cc-pVTZ & cc-pVQZ \\
\hline
\multicolumn{4}{c}{$r_{\ce{N-N}} = r_e$} \\
$\mathcal{C}$   & 0.1567 & 0.1569 & 0.1570 \\
$\mathcal{M}_{1\pi_{u,x} 1\pi_{g,x}}$             & 0.0317 & 0.0318 & 0.0318 \\
$\mathcal{M}_{2\sigma_u 1\pi_u}$          & 0.0046 & 0.0046 & 0.0046 \\
$\mathcal{M}_{2\sigma_g 3\sigma_u}$       & 0.0017 & 0.0017 & 0.0018 \\
$\mathcal{M}_{3\sigma_g 3\sigma_u}$       & 0.0016 & 0.0016 & 0.0016 \\
\hline 
\multicolumn{4}{c}{$r_{\ce{N-N}} = 2r_e$} \\
$\mathcal{C}$   & 2.5025 & 2.4928 & 2.4905 \\
$\mathcal{M}_{1\pi_{u,x} 1\pi_{g,x}}$             & 0.3684 & 0.3681 & 0.3680 \\
$\mathcal{M}_{3\sigma_g 3\sigma_u}$       & 0.2776 & 0.2762 & 0.2760 \\
$\mathcal{M}_{1\pi_{u,x} 1\pi_{u,y}}$           & 0.0120 & 0.0120 & 0.0120 \\
$\mathcal{M}_{1\pi_{u,x} 1\pi_{g,y}}$         & 0.0093 & 0.0091 & 0.0090 \\
\hline

\hline    
\end{tabular}
\label{tab:basis_set}
\end{table}

\begin{figure}[hbt!]
    \includegraphics[width=3in]{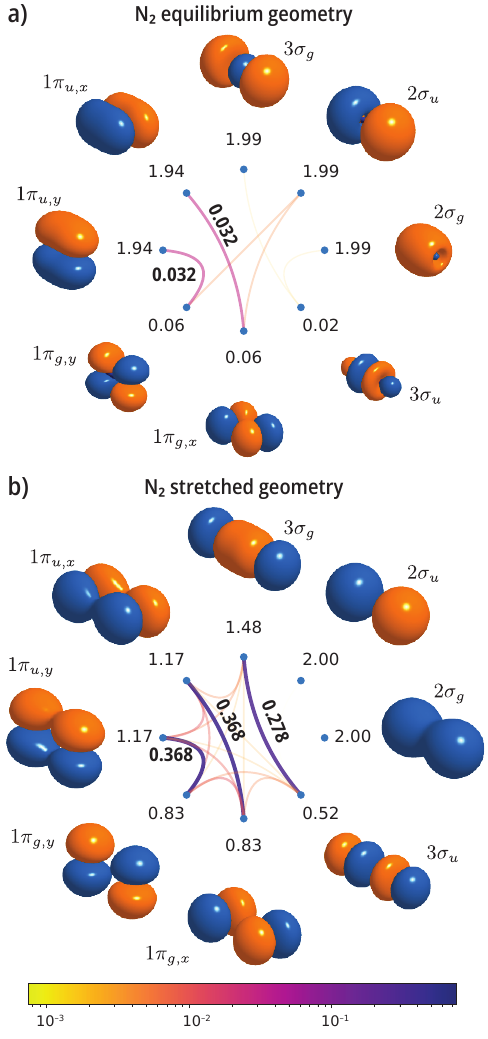}
    \caption{Singlet ground state of the \ce{N2} molecule. Mutual correlation plots computed with a valence active space comprising the 2s and 2p orbitals [CASSCF(10,8)] using a cc-pVDZ basis set at (a) the experimental geometry ($r_e$ = \SI{1.098}{\angstrom})  and (b) a stretched geometry (2$r_e$).}
    \label{fig:n2}    
\end{figure}

Next, we consider a typical use case of the mutual correlation metric: establishing the extent of strong correlations and the corresponding orbitals involved.
To this end, we visualize $\mathcal{M}_{A_P A_Q}$ in mutual correlation plots, following a graphical approach introduced in Ref.~\citenum{Boguslawski.2012.10.1021/jz301319v} to represent orbital mutual information.
These plots show orbitals arranged in a circle, ordered by decreasing occupation number.
The value of the orbital mutual correlation between is displayed with lines that connect orbital pairs, with color and width dependent on the value of $\mathcal{M}_{PQ}$. We find it useful to display $\mathcal{M}_{PQ}$ with a range spanning three orders of magnitude, with colors ranging from yellow (0.00075) to dark violet (0.75). Values that fall below this range are not shown.

To build a sense for how the value of $\mathcal{M}_{PQ}$ connects to chemical concepts, we focus on the \ce{H2} molecule.
In \cref{fig:h2}, we show mutual correlation plots at two geometries ($r_e$ = \SI{0.74144}{\angstrom} and 3$r_e$).
We take \ce{H2} at the equilibrium geometry to represent typical closed-shell molecules with a weakly correlated electronic structure.
This is reflected in the occupation number of the first natural orbital (1.97), which is close to double occupancy.
For this case, the largest value of the orbital mutual correlation is 0.006, which we consider at the low end of the range and indicative of weak electron correlation.
\ce{H2} at stretched geometries is instead a prototypical molecular system with strong electron correlation.
The onset of strong correlation is accompanied by an increase in occupation number for the $1\sigma_u$ orbital (which goes from 0.01 to 0.54).
The maximum value of the orbital mutual correlation is $\mathcal{M}_{1\sigma_g 1\sigma_u}$ = 0.528, signaling strong correlation between the $1\sigma_g/1\sigma_u$ bonding-antibonding orbital pair.
Another important part of the \ce{H2} dissociation curve is the so-called recoupling region, where the Hartree--Fock solution undergoes restricted to unrestricted symmetry breaking (Coulson--Fisher point).
When we examine one point in this region (at 2$r_e$), the maximum orbital mutual correlation takes the value $\mathcal{M}_{1\sigma_g 1\sigma_u}$ = 0.138.

Following this analysis, we propose considering three ranges for the value of $\mathcal{M}_{PQ}$ as follows:
1) 0.75--0.075 (strong correlation), values in this range signal strong mutual correlation, typically associated with breaking a chemical bond. The upper limit is consistent with the maximum value taken by $\mathcal{M}_{PQ}$ for the two-electron model studied in \cref{sec:partitioning}.
2) 0.075--0.0075 (medium correlation), this range captures mid-to-weak electron correlation between bonding/antibonding orbitals that do not display degeneracies.
3) 0.0075--0.0 (weak correlation), values that fall within this range are consistent with very weak electron correlation, as observed in the \ce{H2} $1\sigma_g$/$1\sigma_u$ bonding/antibonding pair.

We caution the reader that this scale is arbitrary.
In this work, we will employ this classification only for the purpose of simplifying the analysis of orbital mutual correlation plots.

\subsection{Analysis of representative molecules}

\begin{figure}[hbt]
    \includegraphics[width=3in]{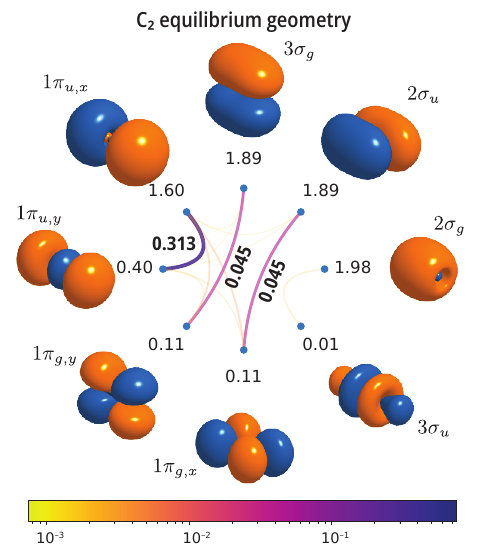}
	
    \caption{Singlet ground state of the \ce{C2} molecule. Mutual correlation plots computed with a valence active space comprising the 2s and 2p orbitals [CASSCF(8,8)] using a cc-pVDZ basis set at the experimental geometry ($r_e$ = \SI{1.2425}{\angstrom}).}
    \label{fig:c2}    
\end{figure}

Our next example considers the triple bond in \ce{N2}, focusing on the valence orbitals.
In \cref{fig:n2}, we show mutual correlation plots for the singlet ground state at the experimental geometry $r_e$ = \SI{1.098}{\angstrom} (a) and 2$r_e$ (b).\cite{Huber.1979.10.1007/978-1-4757-0961-2_2}
At the equilibrium geometry, the most notable group of weakly correlated orbitals are the $1\pi_u$/$1\pi_g$ pairs, with the antibonding combinations being weakly occupied.
At the stretched geometry, a pattern arises, with large values of the mutual correlation among pairs of $3\sigma_g/3\sigma_u$ and $1\pi_u$/$1\pi_g$ orbital pairs.
In this example, correlation is captured by a product of three geminal functions correlating an electron pair in two orbitals.
It is interesting to contrast the equilibrium geometry mutual correlation plot of \ce{N2} with that of \ce{C2}. As shown in \cref{fig:c2}, already at the equilibrium geometry, \ce{C2} shows strong mutual correlation between the $3\sigma_g/3\sigma_u$ orbitals, reflecting the partial occupation and diradical character of this molecule.
In contrast, correlation among the $1\pi_u$/$1\pi_g$ orbital pairs is weak, like in the case of \ce{N2}.

\begin{figure}[hbt!]
    \includegraphics[width=3in]{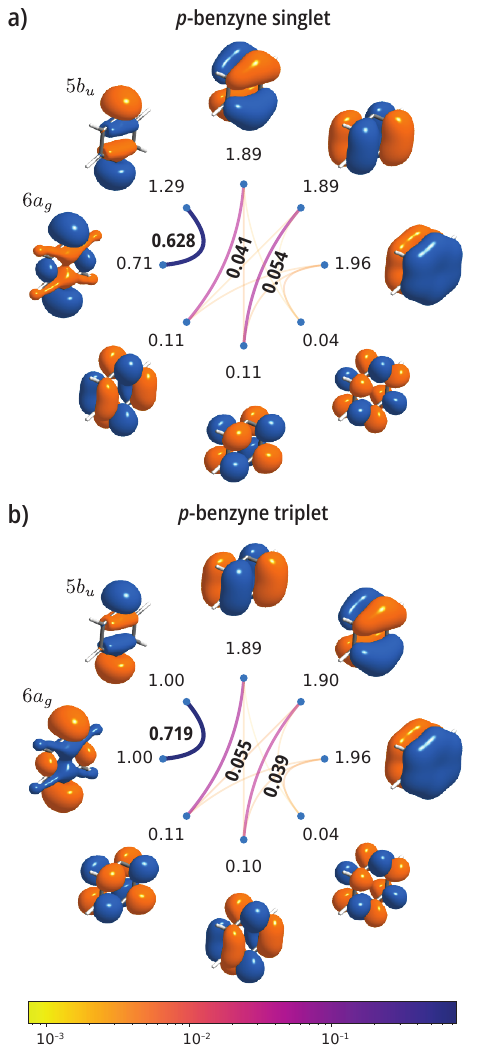}    
    \caption{\textit{p}-benzyne. Mutual correlation plots for the lowest singlet (a) and triplet (b) states computed with an active space comprising the radical $\sigma$ orbitals ($5b_u$, $6a_g$) and six $\pi$ orbitals [CASSCF(8,8)] using a cc-pVDZ basis set. Both states were computed with the geometry for singlet \textit{p}-benzyne ground state taken from Ref.~\citenum{Evangelista.2012.10.1063/1.4718704fix}.}
\label{fig:p-benzyne}
\end{figure}

Finally, we examine the singlet and triplet states of \textit{p}-benzyne.
Both plots shown in \cref{fig:p-benzyne} show the same correlation pattern in both states, with the mutual correlation between the radical orbitals ($5b_u$, $6a_g$) being greater in the triplet state (0.719) than in the singlet state (0.628).
This analysis also reveals that both states are separable into a product of cluster states of the $\sigma$ and $\pi$ orbitals, suggesting an approximate treatment via a restricted or generalized active space.

\subsection{Maximally correlating orbitals}

Having defined orbital mutual correlation, we consider a set of orbitals that localizes the mutual correlation to the smallest number of orbital pairs.
We note that a similar idea was recently proposed in the context of entanglement-based metrics by Liao, Ding, and Schilling\cite{Liao.2024.10.1021/acs.jpclett.4c01105} by defining orbitals that minimize the sum over all orbitals of the one-orbital entanglement.

We consider unitary transformations of the spin orbital basis:
\begin{equation}
\ket{\psi'_p} = \sum_{q} \ket{\psi_q} U_{qp}
\end{equation}
where $U$ is a unitary matrix parameterized by an anti-Hermitian matrix $A$ via the exponential:
\begin{equation}
U = e^{A}
\end{equation}
We restrict our analysis to matrices $U$ that separately  transform spin up and down spin orbitals:
\begin{equation}
U =
\begin{pmatrix}
 U^{\uparrow} & 0 \\
 0 & U^{\downarrow}
 \end{pmatrix},
\end{equation}
and consider restricted transformations with $ U^{\uparrow} = U^{\downarrow}$.
To localize the mutual correlation, we define a cost function given by the sum of the squares of the mutual correlation, generalizing a common approach in orbital localization.\cite{Pipek.1989.10.1063/1.456588,Lehtola.2014.10.1021/ct401016x}
The cost function $\mathcal{L}$ is defined as the sum of the squares of the mutual correlation:
\begin{equation}
\label{eq:cost}
\mathcal{L} = \sum_{P<Q} (\mathcal{M}_{PQ})^2
\end{equation}
which we numerically maximize with respect to the unique entries of the anti-Hermitian matrix $A$.
Due to the nonlinear nature of the cost function $\mathcal{L}$, we expect the existence of multiple local minima.

\begin{figure}[hbt!]
    \includegraphics[width=3in]{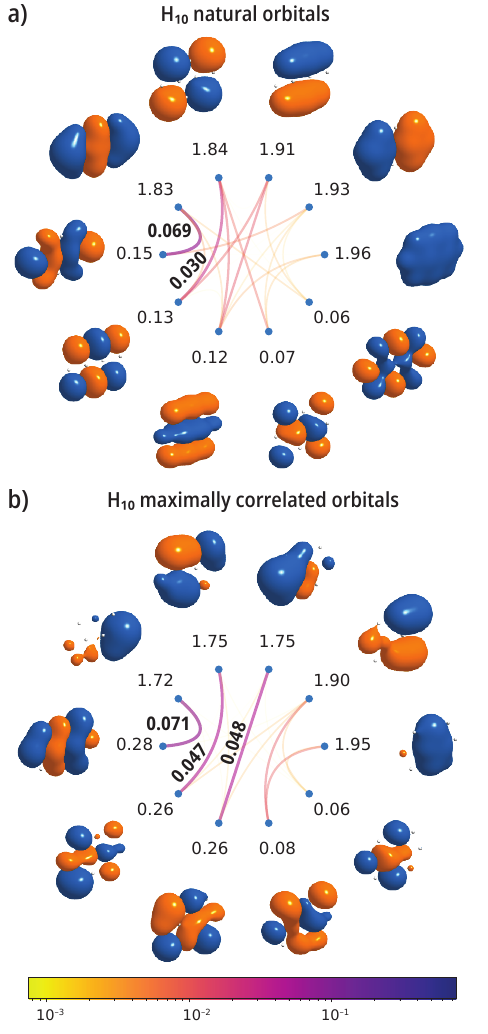}
    \caption{Singlet ground state of a 2D \ce{H10} sheet with nearest neighbor distance $r_\text{H-H}$ = \SI{1.5}{\angstrom}. Mutual correlation plots computed with a valence active space comprising the 1s orbitals [CASSCF(10,10)] using a cc-pVDZ basis set. (a) natural CASSCF orbitals and (b) maximally correlated orbitals.}
    \label{fig:h10_2d_opt}    
\end{figure}

We first consider a system of 10 hydrogen atoms in a planar configuration (triangular pattern) and with a nearest-neighbor distance of \SI{1.5}{\angstrom}.\cite{Stair.2020.10.1063/5.0014928}
For this example, we consider a full valence CASSCF state with ten electrons in ten orbitals.
\Cref{fig:h10_2d_opt} a) shows the mutual correlation plot of \ce{H10} in the natural orbital basis. This plot shows medium-to-high-range orbital mutual correlation for six orbitals with occupation numbers ranging from 1.91 to 0.12.
In the natural basis, the largest value of the mutual correlation is 0.069 and occurs for the nominal HOMO and LUMO pair.
When the cost function \cref{eq:cost} is maximized, we obtain the orbitals shown in \cref{fig:h10_2d_opt} b).
In contrast to natural orbitals, maximally correlated orbitals break the molecular point group symmetry.
In the transformed orbital basis, the occupation numbers deviate more from the doubly occupied (2) or empty (0) case.
The major effect of the transformation, is the localization of the orbital mutual correlation into three disjoint orbital pairs, with $\mathcal{M}_{PQ}$ values equal to 0.071, 0.047, 0.048.

\begin{figure}[hbt!]
    \includegraphics[width=3in]{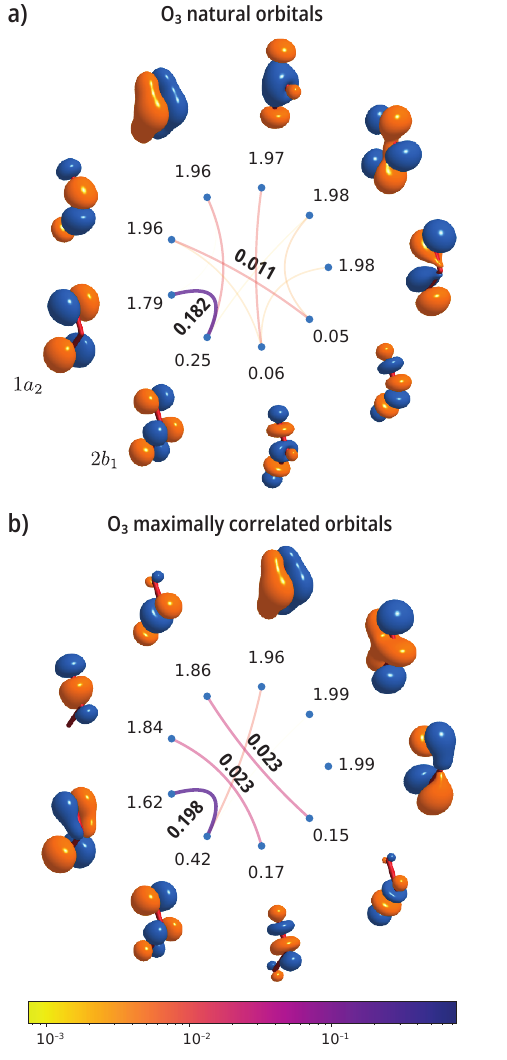}
    \caption{Singlet ground state of the \ce{O3} molecule. Mutual correlation plots computed with a valence active space comprising 12 electrons in 9 orbitals [CASSCF(12,9)] using a cc-pVDZ basis set. (a) natural CASSCF orbitals and (b) maximally correlated orbitals.
    The geometry is taken from Ref.~\citenum{Keller.2015.10.1063/1.4922352}.}
    \label{fig:o3}    
\end{figure}

Our next example considers the ozone molecule, a molecule with partial unpaired electron character. As shown in \Cref{fig:o3} a), in the natural orbital basis, the strongest pair of correlated orbitals consists of the $1a_2$ and the $2b_1$ pair, with a mutual correlation value of 0.182.
Upon localization of the mutual correlation, the orbitals break symmetry and in certain cases spatially localize to atom pairs, forming, for example, \ce{O-O} $\sigma$ and $\sigma^*$ bonds.
The largest value of the mutual correlation increases to 0.198. The $\sigma$ and $\sigma^*$ orbitals localized on each \ce{O-O}  bond form two distinct pairs of orbitals that show modest values of mutual correlation.

Both examples illustrate how localizing the orbital mutual correlation results in a simplified picture of correlation in which orbital pairing emerges naturally.

\subsection{Comparison with Orbital Entropy and Mutual Information}
Lastly, we compare the orbital mutual correlation and mutual information metrics.
The starting point of the information-theoretic metrics is the partitioning of a system into two disjoint subsystems $A$ and $B$ and the corresponding decomposition of a state in their respective bases ($\{\ket{a}\}$ and $\{\ket{b}\}$):
\begin{equation}
\ket{\Psi} = \sum_{ab} c_{ab} \ket{a} \otimes \ket{b}
\end{equation}
In the case of the one-orbital entropy, one defines the subsystem $A$ to be a single spatial orbital $\phi_P$, with corresponding basis spanned by the one-orbital Fock space ($\mathbb{F}_1$):
\begin{equation}
\mathbb{F}_1 = \{ \ket{0},\ket{\uparrow},\ket{\downarrow},\ket{\uparrow\downarrow} \}
\end{equation}
Next, elements of the one-orbital reduced density matrix ($\rho_P$, a $4 \times 4$ matrix) are defined as a projection of the state $\Psi$ onto all outer products $\ket{a}\bra{a'}$ of orbital $\phi_P$, tracing out the remaining degrees of freedom:
\begin{equation}
\label{eq:oneorbrdm}
(\rho_{P})_{aa'} = \mathrm{Tr}_B \left(\ket{a}\bra{a'}\Psi\rangle\langle\Psi|\right) \quad
\ket{a}, \ket{a'} \in \mathbb{F}_1
\end{equation}
The one-orbital entropy\cite{Legeza.2003.10.1103/physrevb.68.195116fix} is then defined as the von Neumann entropy of $\rho_P$:
\begin{equation}
\label{eq:oneorbentropy}
s_P = - \mathrm{Tr}_A \left(\rho_P \ln \rho_P \right)
\end{equation}
As shown in Ref.~\citenum{Rissler.2006.10.1016/j.chemphys.2005.10.018}, the elements of $\rho_P$ may be expressed in term of elements of the 1- and 2-RDMs:
\begin{equation}
\begin{split}
(\rho_{P})_{11} & = 1 - \gamma^{P\uparrow}_{P\uparrow}
- \gamma^{P\downarrow}_{P\downarrow}
+
\gamma^{P\uparrow P\downarrow}_{P\uparrow P\downarrow}\\
(\rho_{P})_{22} & = \gamma^{P\uparrow}_{P\uparrow}
-\gamma^{P\uparrow P\downarrow}_{P\uparrow P\downarrow}\\
(\rho_{P})_{33} & = \gamma^{P\downarrow}_{P\downarrow}
-\gamma^{P\uparrow P\downarrow}_{P\uparrow P\downarrow}\\ 
(\rho_{P})_{44} & = \gamma^{P\uparrow P\downarrow}_{P\uparrow P\downarrow}
\end{split}
\end{equation}

The two-orbital entropy is defined similarly, starting by introducing a basis for the Fock space of two orbitals and consisting of 16 elements:
\begin{equation}
\mathbb{F}_2 = \{ \ket{00},\ket{0\uparrow},\ket{0\downarrow},\ket{0\uparrow\downarrow},
\ket{\uparrow 0},\ldots \}
\end{equation}
For each orbital pair, a two-orbital reduced density matrix ($\rho_{PQ}$) for each pair of orbitals $\phi_P$ and $\phi_Q$ is defined in analogy to \cref{eq:oneorbrdm}.
Like in the case of the one-orbital reduced density matrix, $\rho_{PQ}$ is expressible in terms of RDMs, but with rank up to 4.
The von Neumann entropy of $\rho_{PQ}$ defines the two-orbital entropy:
\begin{equation}
s_{PQ} =
- \mathrm{Tr}\left(
\rho_{PQ} \ln \rho_{PQ}
\right)
\end{equation}
which represents a measure of entanglement of the subsystem spanned by $\phi_P$ and $\phi_Q$ with the rest of the orbitals.
For each orbital pair $\phi_P$ and $\phi_Q$, the mutual information is defined as:\cite{Rissler.2006.10.1016/j.chemphys.2005.10.018}
\begin{equation}
I_{PQ} =
\frac{1}{2} (s_{P} + s_{Q} - s_{PQ}) (1 - \delta_{PQ})
\end{equation}
This quantity is positive and measures the entanglement between orbitals $\phi_P$ and $\phi_Q$.

\begin{figure}[hbt!]
    \includegraphics[width=3in]{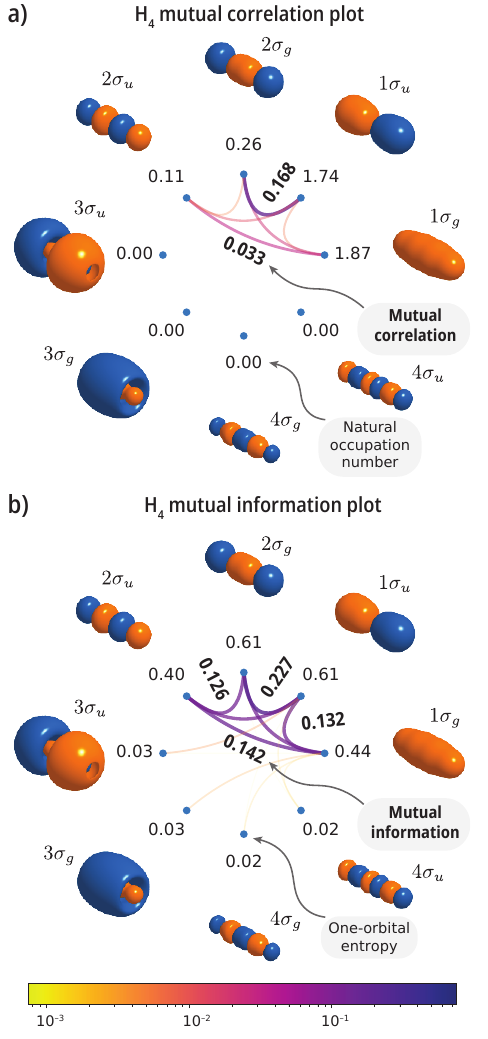}
    \caption{Singlet ground state of the \ce{H4} model system. Mutual correlation and information plots computed with a double valence active space [CASSCF(4,8)] using a cc-pVDZ basis set. Comparison of (a) orbital mutual correlation with (b) orbital mutual information.}
    \label{fig:h4}    
\end{figure}

To compare metrics of correlation, we consider a linear \ce{H4} toy model with nearest-neighbor distance set to \SI{1.5}{\angstrom}.
For this system, we perform a CASSCF computation using a double valence active space comprising an equal number of orbitals belonging to the $A_g$ and $B_{1u}$ irreducible representations.
Due to this choice of active space, we expect to observe a combination of weak and strong correlation effects.
In \cref{fig:h4} we compare metrics based on RDMs (natural occupation numbers and mutual correlation) with their entropy-based counterparts (one-orbital entropy and mutual information).
As evident from \cref{eq:oneorbentropy,eq:oneorbrdm} and the definition of the orbital occupation number ($n_P = \gamma^{P\uparrow}_{P\uparrow}
+ \gamma^{P\downarrow}_{P\downarrow}$), $s_P$ and $n_P$ are related (occupation number appears in the definition of $\rho_P$).
However, these two metrics quantify different aspects of orbital occupancy.
While $n_P$ measures total orbital occupation, $s_P$ quantifies fluctuations in the occupation numbers.
Both one-electron metrics signal noticeable fluctuations in orbital population, as reflected by deviations from occupation 0 or 2, and nonzero one-orbital entropies.
The occupation numbers, however, do not show deviations as extreme as in \textit{p}-benzyne, where they reach values of 1.
This is also the case for the one-orbital entropy, which only reaches values up to 0.61, which should be compared to its upper bound ($\ln 4 \approx 1.386$), achieved when all Fock states of an orbital are equally probable.

The two-orbital measures reflect the fact that the orbitals $1\sigma_g$ through $2\sigma_u$ participate in producing a strongly correlated/entangled ground state.
Although the values of these two metrics fall within similar ranges, the largest mutual correlation value (0.168) emphasizes strong correlation between orbitals $1\sigma_u$ and $2\sigma_g$ and weaker correlation among the remaining orbitals.
Mutual information also identifies $1\sigma_u$ and $2\sigma_g$ as the most strongly entangled orbital pair; however, it assigns a somewhat comparable value to the other interactions among orbital $1\sigma_g$ through $2\sigma_u$.
This comparison highlights the fact that both the Frobenius norm and entropy-based metrics yield the same qualitative picture (the first low-lying orbitals participate in strong correlation/entanglement). However, it also shows that they quantify statistical fluctuations in the orbital occupations differently.

\section{Conclusion}
\label{sec:conclusion}

In this work, we have considered correlation metrics based on the Frobenius norm of the 2-body reduced density matrix cumulant.
From this quantity, we define mutual correlation, a simple and convenient way to quantify pairwise correlation among disjoint subsystems of a many-body system.
We have shown how to systematically decompose the total correlation arising from a general partitioning of the orbitals and defined an orbital mutual correlation, which quantifies correlations due to pairs of orbitals.

We calibrated and examined mutual correlation plots for a variety of prototypical molecules with weakly and strongly correlated electronic structures.
Our examples show that orbital mutual correlation is useful in identifying orbital interactions responsible for strong electron correlation, as well as discerning a separable cluster structure.
A striking result of our analysis, both in the examples presented here and for several other systems examined but not discussed in this work, is the recurring observation that the ground state of molecular systems is dominated by disjoint pairs of correlated orbitals, suggesting that they can be approximated as a product of geminals.
This finding parallels observations based on entanglement-based metrics.\cite{Ding.2023.10.1088/2058-9565/ad00d9}
In our experience, states with more than two mutually correlated orbitals are typically found only as excited-state solutions.
We further consider unitary transformations that localize the orbital mutual correlation to the fewest number of pairs, defining a new set of maximally correlated orbitals.
These maximally correlated orbitals are found to further simplify the analysis of a state, however, at the cost of sacrificing orbital symmetries.
Lastly, we compared metrics based on RDMs with measures of orbital entanglement, identifying similarities.

This study suggests that mutual correlation is an alternative to measures of entanglement that can provide qualitative and quantitative insights into the structure of many-body states.
Since the mutual correlation (either in the general partitioning or the special case of orbitals) depends only on the 2-cumulant, this quantity is, in principle, easily accessible to a variety of quantum many-body methods that provide the 2-RDM, including determinant-based Monte Carlo, configuration interaction, perturbation theory, Green's function, coupled cluster theory, DMRG, and RDM methods.
In contrast, orbital mutual information requires up to the 4-RDM, which limits its application mostly to DMRG and configuration interaction computations.

An interesting alternative left for future studies is to formulate a low-cost approach to quantify entanglement that complements the proposed approach and relies solely on the 1- and 2-RDMs.
This is possible by departing from the orbital-based partitioning\cite{Legeza.2003.10.1103/physrevb.68.195116fix,Rissler.2006.10.1016/j.chemphys.2005.10.018} and instead considering entanglement metrics based on \textit{spin orbital} reduced density matrices.
Other interesting applications of mutual correlation include tracking correlation effects in time-dependent processes, and considering partitions that extend beyond single orbitals to groups of orbitals, enabling the quantification of mutual correlation in interactions of molecular fragments and their excited states.
Lastly, we note that the approach used here to decompose the total correlation, as measured by $\| \boldsymbol\lambda_2\|_\mathrm{F}^2$, from its spin orbital expression is general and can be applied to other metrics of total correlation, as well as extended to observables such as energy and spin.

\begin{acknowledgements}
This work was supported by the U.S. National Science Foundation under Award No. CHE-2312105. The author gratefully acknowledges Dr. Harper Grimsley and Dr. Jonathon Misiewicz for their valuable feedback on the manuscript.
\end{acknowledgements}

\providecommand{\latin}[1]{#1}
\makeatletter
\providecommand{\doi}
  {\begingroup\let\do\@makeother\dospecials
  \catcode`\{=1 \catcode`\}=2 \doi@aux}
\providecommand{\doi@aux}[1]{\endgroup\texttt{#1}}
\makeatother
\providecommand*\mcitethebibliography{\thebibliography}
\csname @ifundefined\endcsname{endmcitethebibliography}
  {\let\endmcitethebibliography\endthebibliography}{}

\end{document}